# Computational Biomechanics, Stochastic Motion and Team Sports

E. Grimpampi[1], A. Pasculli[2] and A. Sacripanti[1,3]
[1] Facoltà di Medicina e Chirurgia,
University of Rome "Tor Vergata", Italy
[2] Facoltà di Scienze MM.FF.NN.,
University G. D'Annunzio, Chieti.Pescara, Italy
[3] Dipartimento Tecnologie della Fisica e Nuovi Materiali (FIM),
ENEA- Italy

## Abstract

In this paper we put the basis for a mathematical theory of competition in situation sports, such as dual sports and team sports. It is shown that in dual contest sports, the motion of the centre of mass of a couple of athletes is well described, with a good approximation, by Classical Brownian Motion. In contrast, the problem of the motion in team sports, like soccer, football, basketball, water polo, and so on, seems more complex and it would be better to be modelled by a special class of Brownian Motion, the well known Active Brownian Motion, with internal energy depot. In this paper a special equation is proposed for the first time, describing the athletes motion in team sports game and a numerical simulation of the trajectories. The motion paths, obtained from the computational approach, are validated using experimental data of actual games, obtained from motion analysis systems.

**Keywords:** sport biomechanics, stochastic modelling, Brownian motion, complex many-particle systems, dual sports, self organization

# 1 Introduction

It is well known that the evolution of the self organizing complex organic systems is described by their non linear evolution in time.

If we observe at microscopic and mesoscopic scale, with specific attention to the human body, we can find that all the inside physiological self organized complex structures such as the DNA, the coronary artery tree, the Purkinjie cells in cerebellum, the small intestine and others, exhibit the property of self-affinity, that is the natural form of the well known geometrical property of self similarity.

Self similarity is a well known property of fractals structures, and we can find it in the whole human body, in their static, kinematics and dynamics forms [1].

The connection among these different aspects of the human body as a complex system is the generalized Brownian Motion in its every known formulation: classic, fractional, active and so on.

It can be shown that starting from fractals and finishing with multifractal aspects of the human physiological complex systems or response, Brownian Dynamics is one of the basic modelling of biological systems.

But if we study in deep the evolution of the macroscopic complex systems in time connected to the human body, starting from the motion of centre of mass in standing still position to gait, the Brownian Motion shows its ubiquitous presence in the description of these phenomena, as in the case of the Fractional Langevin equation which describes the variability of the stride interval during walking [2, 3].

More surprising, if we study the time evolution of macroscopic self organizing complex systems consisting in more than one human bodies, once more Brownian Dynamics are present.

The above mentioned applications of Brownian Motion are well known in the case of crowd queues or evacuation problems, but it hasn't been applied in the specific fields of sport Biomechanics, such as the so called "Situation Sports" [4]. The "Situation Sports" are identified as *"sports in which the independence of simultaneous actions is not applicable for studying the athlete's motion in competition; in these specific field it is better to use a more sophisticated approach using statistical physics and the chaos theory"*. Among these sports, a number of contact sports can be identified (both dual and team sports), which exhibit a most complex motion during competition. This is the case of fighting sports (dual sports) and the more well known team sports, such as soccer, basketball, water polo, football, hockey, etc.

As it has already been shown, the Classical Brownian Motion of a "simple" particle [5] describes in good approximation the motion of the centre of mass of a couple of athletes in competition, in the case of dual sports such as judo, karate, boxing, wrestling, etc.

The problem of the motion in team sports is a little bit more complex. First of all, the theoretical approaches can be broadly divided into two categories:

a) "Individual-based"; and
b) "Team-based".

The "Individual-based" models explicitly describe the dynamics of the individual elements. As the "microscopic" models of matter are formulated in terms of molecular constituents, also the "Individual-based" models of transport are developed in terms of the constituent elements.

In contrast, the individual elements in the "Team-based" models do not appear explicitly but one considers only the population densities (i.e., number of athletes per unit area or per unit volume) [6].

The space-temporal organization of the athletes shows collective properties, which are determined by the response of individuals to their local environment and by the local interactions among the athletes of a different team. Therefore, in order to gain a deep understanding of the collective phenomena, it is essential to investigate the relationship between these two levels of organization [7].

In this paper an "Individual based" approach is explored, based on the Active Brownian Motion, and consequently, the single trajectory obtained is studied. Also the "Team based" approach is discussed and some general information from other fields of physics are seeked, with a well known synergetic approach

# 2    Situation sports

In the field of Sport Biomechanics, a sports classification is necessary to be established in order to study the performance of the athlete. This classification can be manifold: in physiology for example, the classification can be performed in function of the performance energy expenditure.

In Sport Biomechanics the most functional classification is the biomechanical one [4], which is established in function of few basic movements during the performance. This classification allows to pursue the most basic complex movements that must be measured by specific scientific discipline, both as a whole or in step sizes.

In alternative these most basic movements could be the goal of an expert group such as of a sport physiologist, a neurologist, a sports biomechanics specialist, an engineer, a trainer, or a technician, for improving the athletes performance. The aim of this classification is to allow to properly find what kind of a specific observational approach must be applied to solve the problem by both qualitative and quantitative mechanical or mathematical models.

This classification allows to classify all the sports in four big categories [4]:
1. **Cyclic Sports**. All the sports in which the basic movement is repeated continuously in time, like gait, running, marathon, cycling, swimming, rowing, etc.
2. **Acyclic Sports**. All the sports in which the basic movement is applied only once during the performance, like: discus, shot put, hammer throw, pole vault, high jump, long jump, triple jump, ski jump, javelin throw, etc.
3. **Alternate-Cycling Sport**. All the sports in which two (o more) basic movements are applied, alternatively in time, like 110 hurdles, 400 hurdles, steep-chase, golf., etc.
4. **Situation Sport**. All the sports in presence of an adversary; these sports, can be divided in two classes (without and with contact) and each class in two sub classes (dual sports and team sports).

Situation dual sports without contact include tennis and ping pong, while situation team sports without contact include volleyball and beach volley.

Situation dual sports with contact are all the fighting sports such as: judo, boxing, wrestling, karate, kick boxing, Wu Shu, Exrimia, etc., while situation team sports with contact are, among other, soccer, basketball, football, water polo, hockey.

The situation sports are sports where it is not possible to find a repeatable motion pattern for each specific match. For each match the corresponding motion is a random process and there are no basic specific movements during the motion, but it is possible to find these repeatable movements only during the interaction among athletes. In fact the correct way to analyze such macro phenomena is to study them in two steps: motion and interaction with basic repeatable movements.

It is remarkable that motion in competitions for each class of these sports could be associated with the well known Brownian Motion. In fact, if we consider for each sport the basic motion pattern of a big number of matches from a statistical point of view, like in the classic Gaussian approach, it is straightforward that the motion belongs to the classes of Brownian Motion as it is shown in the next paragraph.

## 2.1 Dual Sports

The first step in the performed biomechanical modelling is to describe the problem of the competition in contest sport. This problem was faced and solved by one of the authors sixteen years ago [5], both in theoretical and experimentally. The main results of this very old research are summarised below.

In contest sports (as is the case for situation sports with contact, Figure 1), it is often impossible to describe or to understand the duel, because each contest is different from the other; however, if we consider each contest as the single representative of one set of infinite members, the problem can be approached by means statistical mechanics methods.

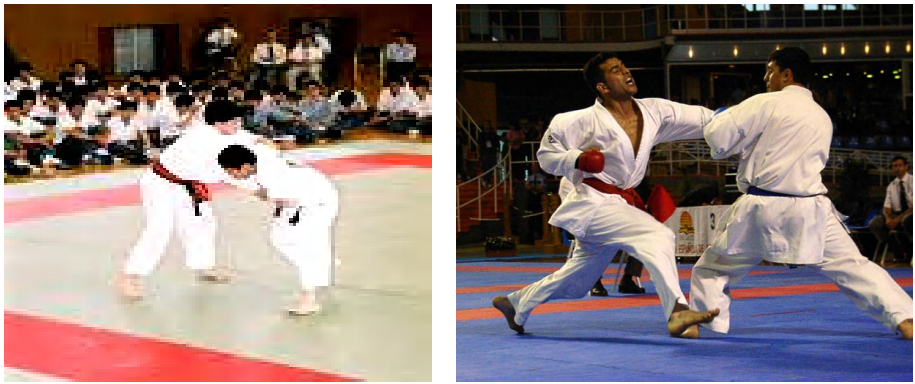

Figure 1: Situation dual sports with contact.

In modelling performed, the simplification is rather extreme, but the physical and biomechanical meaning is preserved. If we consider the couple of athletes as a single system, then the motion of the centre of mass system is definite by a push pull random forces, which are directly connected to the friction forces between feet and mat.

This is the biomechanical base of the known judo paradox that state: *"the most important part of the grip is the feet position"*,

Then the push pull random forces could be described as:

$$\varphi(t) = u \sum_j \delta(t - t_j) \tag{1}$$

in which $u = m\Delta v$.

The system as a whole is isolated, with no external forces other than the random push-pull forces by the friction force, and the general equation of motion of the athletes fighting is a Langevin type equation [5]. It is thus possible to write:

$$F = \dot{v} = -\frac{\mu}{m}v + \frac{u}{m}\sum_j (\pm 1)_j \delta(t-t_j) = F_a + F' \qquad (2)$$

Along the line of Einstein, the statistical relation for the stationary kinetic energy is [8]:

$$E_c = \frac{1}{2}m\langle v^2 \rangle = \frac{m^2}{4\mu}C \qquad (3)$$

where, with reference to the athletes, one of the authors proposed the following equation [8], in order to relate the mechanical with the physiological data:

$$E_c = \frac{1}{2}m\langle v^2 \rangle = \rho \overline{V}_{0_2} \approx \frac{1}{5}\overline{V}_{0_2} \qquad (4)$$

and

$$C = \frac{4}{5}\frac{\mu}{m}\overline{V}_{0_2} \qquad (5)$$

In this Individual–fight based model, every contest is a Markov process, and all the contests are individually independent. Considering the well known work of Smoluchovski [9] on the Brownian Motion, the "Physics" that produce the random evolution of the contest allows us to obtain the basic probability of this Markov process.

As a consequence, in the case of dual sports, it is possible to obtain from the transition probability Q, the solutions of the conditional probability which provides, at an infinite time limit, the probability to find an athlete between $x$ and $x + dx$ at time t; in mathematical form it is possible to write:

$$Q(k,m) = \frac{1}{2}\delta(m, k-1) + \frac{1}{2}\delta(m, k+1) \qquad (6)$$

The solution to the above equation is:

$$P(n|m,s) = \frac{s!}{\left(\frac{v+s}{2}\right)!\left(\frac{v-s}{2}\right)!}\left(\frac{1}{2}\right)^s \qquad (7)$$

The experimental proof of this model can be found in one Japanese study [10], in the 1971 world Judo championship.

In Figure 2 we can see the summation of motion patterns of 1, 2, 7, and 12 judo fights; it is easy to understand that the random fluctuation doesn't have a preferential direction over the time. In mathematical terms this means that:

$$\langle F' \rangle = 0 \qquad (8)$$

Therefore, it is possible to assert that the motion of the centre of mass of the systems is Brownian.

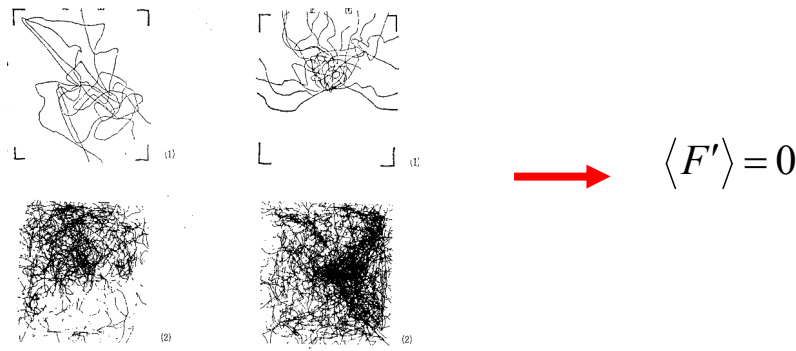

Figure 2: Motion patterns of judo fighters [10].

## 2.2 Team Sports as Self Organizing Complex many-particle Systems

A good model for describing Team sports, from the "Team based" point of view, is to consider them as self organizing complex systems. From a modelling point of view of the main argument, the team sport contest can be described as a cyclic continuum in a time Markov process, having the property of self organization.

An interesting way to approach complex systems derives from a special view of the Random walks. It is based on incorporating the complexity of the system in the Random walk itself, introducing memory in it through fractional differences [11]. If we look more carefully at the dynamic aspect of the process it is possible to write a generalized fractional Langevin equation and introduce the Fractional Brownian Motion. In mathematical form it is possible to write:

$$D_t^\alpha [X(t)] - \frac{X(0)}{\Gamma(1-\alpha)} t^{-\alpha} = \xi(t) \qquad (9)$$

where the first term is a fractional derivative, the second is connected to the initial condition of the process, and the third is always the random force acting on the particle.

In this case is important to identify the mean square displacement of the particle:

$$\left\langle [X(t)-X(0)]^2 \right\rangle = \frac{\langle \xi^2 \rangle}{(2\alpha-1)\Gamma(\alpha)^2} t^{2\alpha-1} \propto t^{2H} \qquad (10)$$

From the above expression it is possible to recognize, that an anomalous diffusion process is present, identified by the *H* parameter usually called *Hurst parameter*; in particular this parameter is time independent, and it describes the fractional Brownian motion with anti-correlated samples for *0<H<½* and with correlated samples for *½<H<1*. If *H* is equal to ½ we have a Brownian Motion.

In general *H* can be a function of time; in recent times this important extension is called *multi-fractional Brownian motion* [12].

This important generalization derives from a certain situation occurring either in the field of turbulence [13] or from Biomechanics [14] where a more flexible model is needed, necessary both to locally control the dependence of the structure and to allow the path to vary regularly in time (Figure 3).

   a. *Pure Brownian motion: next step is uncorrelated with previous step H=0.5*
   b. *Persistent Fractional Brownian motion: each step is positively correlated with previous step H<0.5*
   c. *Anti-Persistent Fractional Brownian motion: each step is negatively correlated with previous step H> 0.5*

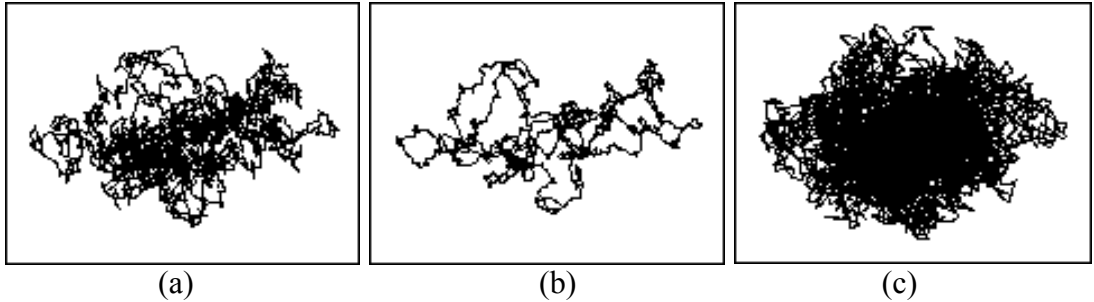

(a)          (b)          (c)

Figure 3: Brownian motion patterns

In the Individual-based point of view, it is possible to equally account for the complexity and the self organization of a team in several different ways, the two most useful being in our point of view the Active Brownian Motion and the introduction of the "Social Force".

## 2.3 Modelling of Team Sports

The theoretical approach to the team sports, can be broadly divided in two categories:
- "Individual-based"; and
- "Team-based".

The individual-based models describe the dynamics of the individual elements explicitly.

Just as "microscopic" models of matter, the "Individual-based" models are developed in terms of the constituent elements.

In contrast, the individual elements in the Team-based models do not appear explicitly but only the population densities can be taken into account (i.e., the number of athletes per unit area or per unit volume).

In the optics of the "Individual-based" models, the first model [5], relative to the dual situation sports, showed that the motion of the centre mass of couple of athletes systems is a classical Brownian Motion. That means that there is not a preferential direction in their motion patters.

In the team sport approach as "Individual based", the description of the situation is completely different. There is a special preferential direction in the motion pattern and they are self organized. Also in team sports, every single athlete is not in stable equilibrium as in the case of the system of a couple of athletes.

In order to describe this case it is necessary to adopt a more complex model for the athlete's motion, like the Active Nonlinear Brownian Motion proposed by Ebeling and Schweitzer [15]. It is also necessary to introduce the "Social force" [16] for both of mutual interaction and self-organization.

In this modelling it is possible to consider and take into account the oxygen uptake from the environment and the potential interaction against the other adversary team members in relation to the self organization of the whole team.

As for the Equation (4), it is possible to write for the internal energy variation:

$$\frac{dE(t)}{dt} = \dot{V}_{0_2}(t) - \eta(v^2) K v^2 E(t) \tag{11}$$

The above indicates that the internal energy varies in terms of input-output flux, mainly the oxygen uptake that converts himself in external kinetic energy.

If we adopt the hypothesis that the internal energy $E(t)$ varies slowly, Equation (11) can be simplified on the basis of the following assumptions:

$$\frac{dE(t)}{dt} \approx 0$$
$$\dot{V}_{0_2}(t) \cong \overline{V}_{0_2}(t) \tag{12}$$

As a result it is possible to obtain the special value for the energy $E_0(t)$:

$$E_0 = \frac{\overline{V}_{0_2}}{\eta K v^2} \tag{13}$$

The equation includes a term $kE_0 v$ adopted by Ebeling [17] and the nonlinear friction coefficient. In this case will be:

$$\gamma_v = \gamma_0 + E_0 \equiv \gamma_0 - \frac{k\overline{V}_{0_2}}{\eta k v_0^2} = \gamma_0 - \frac{\overline{V}_{0_2}}{\eta v_0^2} \tag{14}$$

Considering the potential interaction against the adversary, the collision or the avoidance, the second model can be defined. On the basis of the Ebeling, Schweitzer and Helbing equations, the following Langevin type equation for a nonlinear Brownian Motion proposed in the Sacripanti's second model, is in compact form:

$$\begin{aligned}ma &= -\gamma_v v + F_{acc} + \left[\sum F_1 + \sum F_2\right] + \left(\sqrt{2D(v)}\right)u\sum(\pm 1)\delta(t-t') = \\ &= -\gamma_v v + F_{acc} + \left[\sum F_1 + \sum F_2\right] + F'\end{aligned} \tag{15}$$

If we account in explicit form the nonlinear motion, the oxygen uptake, the kinetic energy from uptake and potential mechanical interaction like collision and avoidance manoeuvres, it is possible to write:

$$\begin{aligned}ma &= \left(\gamma_0 - \frac{\overline{V}_{0_2}}{\eta v^2}\right)v(r,t) + \frac{m}{t}\left[v^0 e(t) - v_1\right] + \eta k\left(r_{1,2} - d_{1,2}\right)N_{1,2} + \\ &+ A_{1,2}N_{1,2}e^{\left(\frac{r_{1,2}-d_{1,2}}{B}\right)}\left[\lambda_1 - (1-\lambda_1)\frac{1+\cos\varphi_{1,2}}{2}\right] + \left[2D(v)\right]^{\frac{1}{2}}u\sum_j(\pm 1)\delta(t-t_j)\end{aligned} \tag{16}$$

Where:
- $\left(\gamma_0 - \overline{V}_{0_2}/\eta v^2\right)$ indicates a nonlinear friction coefficient depending from the athlete's oxygen uptake;
- $\left[v^0 e(t) - v_1\right]/t$ indicates the acceleration term with the desired velocity;
- $\eta k(r_{1,2} - d_{1,2})N_{1,2}$ represent the pushing force; and
- $A_{1,2}N_{1,2}e^{\left(\frac{r_{1,2}-d_{1,2}}{B}\right)}\left[\lambda_1 - (1-\lambda_1)\frac{1+\cos\varphi_{1,2}}{2}\right]$ represents the social repulsion force between the athletes.

The original parameters were modified in a recent study [18]. The specific preferred direction in motion patterns of the team sports, can be modelled by a proposed solution model [19]. With a special modification made by one of the authors it is possible to model the basic probability of this Markov process (the game) in function of the special attack strategy adopted.

In fact, in the case of team sports it is possible to write the transition probability $Q$ in function of the attack strategy $\alpha$. The $\alpha$ parameter may vary from 1 to 5, with the following meanings:
1 = lightning attack; 2 = making deep passes; 3 = manoeuvring; 4 = attack by horizontal passes; 5 = melina.

The solution of the conditional probability $P$ are connected to the limit of mean value in time for finding the athlete between $x$ and $x + dx$ at time $t$, in formulas:

$$Q(k,m) = \frac{R^\alpha + k}{2R^\alpha}\delta(m, k-1) + \frac{R^\alpha - k}{2R^\alpha}\delta(m, k+1)$$
$$with\ -1 \leq \alpha \leq 5 \tag{17}$$

$$\langle m(s)\rangle_{av} = \sum_m mP(n[m,s]) = \left(1 - \frac{1}{R^\alpha}\right)\langle m(s-1)\rangle_{av}$$

On the same topic, a very interesting study for the diffusion coefficient of Active Brownian motion [20], provides the very important result that the Diffusion coefficient is decreasing in function of the noise intensity (Figure 4). It is also possible to approximate it as function of the square velocity of the athlete divided the Kramer escape rate:

$$D_{eff} \approx \frac{v_0^2 \pi}{Q\sqrt{U''(v_0)|U''(0)|}} e^{\Delta U} = \frac{v_0^2}{2r_k} \tag{18}$$

where $\Delta U$ is equal to:

$$\Delta U = \frac{m\gamma_0\left[1 + \overline{V}(\ln \overline{V} - 1)\right]}{2Qe} \square\ 1 \tag{19}$$

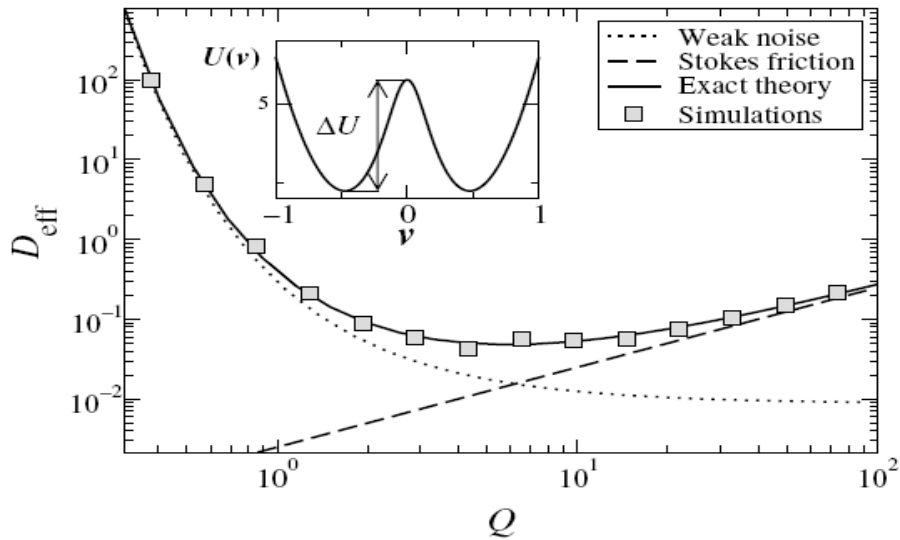

Figure 4: Diffusion coefficient evolution for a nonlinear Active Brownian Motion (adapted from [20])

In Figure 5 it is possible to observe that, despite the preferential direction present in each motion pattern, from a statistical point of view (summation of several motion patters from several games), also in team games, the random fluctuation does not have a preferential direction over the time, implying that $\langle F'\rangle = 0$.

As a consequence, the global motion in this case is Brownian as well.

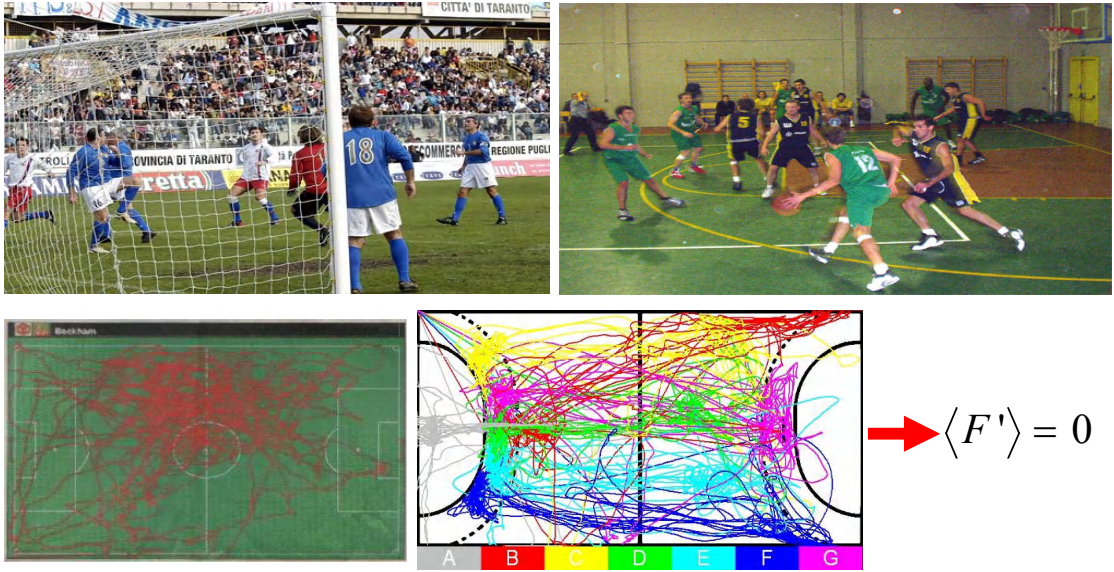

Figure 5: Motion patterns in team games

The space-temporal organization of the athletes are characterised by emergent collective properties, that are determined by the response of the individuals to their local environments and the local interactions among the athletes of different team. Therefore, in order to gain a deep understanding of the collective phenomena, it is essential to investigate the linkages between these two levels of organization.

## 2.4 Trajectories results

The study of the trajectories is a very interesting field in the Brownian Dynamics; to this day in the scientific literature two main approaches are encountered:
  a) one approach is proposed by the Kozlov, Pitman and Yor theory (*KPY theory*) [21] and is useful in analyzing the location of the trajectory on a specific surface.
  b) a second one connected to the inverse dynamics of the trajectory, trying to obtain from his study the potential that can produce the analyzed trajectory, proposed among others by Brillinger [22].

In the first approach, the trajectory of the ball in a soccer game is modelled by the Brownian motion on a cylinder, subject to elastic reflections at the boundary points (as proposed in KPY theory). The score is then the number of windings of the trajectory around the cylinder.

Later, it was considered in addition a generalization of this model due to Baryshnikov [23] to higher genus, proving asymptotic normality of the score and deriving the covariance matrix. Ambiguity of the short paths system contributes just a bounded term to the score( see for example [24]).

In the second and more modern approach the basic data are points in the plane, successively joined by straight lines. The resulting figure represents the trajectory of the moving soccer ball. Defined Brownian in the *KPY theory* and in this work.

The approach of this study is to develop a useful potential function, a concept arising from physics and engineering. In particular, the potential function leads to a regression model that may be fit directly by linear least squares. The resulting potential function may be used for simple description, summary, comparison, simulation, prediction, model appraisal, bootstrapping, and employed for estimating quantities of interest.

Figure 6 shows the application of this inverse physical approach with the ball trajectory and the deduction of the potential function.

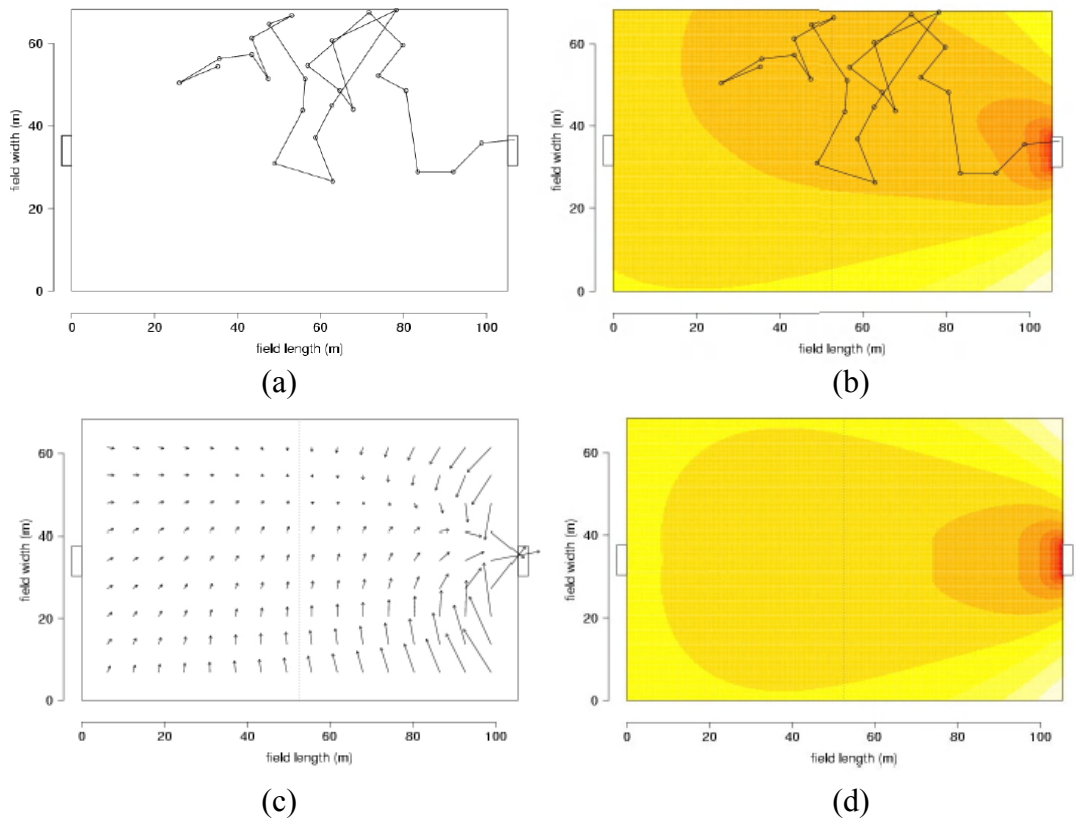

Figure 6: The ball trajectory and the inverse physical approach [24]

## 2.5 Numerical Evaluation

Equation (16) can be expressed in the following vector form where all the parameters are related to a "virtual" player:

$$m\frac{d\mathbf{v}}{dt} = -\beta\mathbf{v} - m\frac{(\mathbf{v}-\mathbf{v}_a)}{\tau} + \mathbf{P} - \mathbf{A}e^{-|x|/b} + \mathbf{L} \qquad (20)$$

In the above m is the mass, v is the vector velocity, $\tau$ is the relaxation time necessary to reach the "target velocity" $v_a$, **P** is the pushing force, $Ae^{-|x|/b}$ is a global term related to the border line distance, the adversaries vicinity and the adopted strategy and **L** is the random Langevin force. It is made the hypothesis that the player has many "interactions" due to tackles, strategy changing, adversary contact and so on. Between each interaction it is assumed that he follows a straight line, while the random Langevin force **L** is supposed to influence only the trajectory direction after each interaction.

Following the abovementioned hypothesis, Equation (20) is solved along a chosen direction considering the following scalar equation:

$$m\frac{dv}{dt} = -\left(\beta + \frac{m}{\tau}\right)v + \left(\frac{mv_a}{\tau} + P - Ae^{-|x|/b}\right) \qquad (21)$$

whose solution is simply:

$$\Delta s = \frac{c}{a}\Delta t + \left(\frac{v_0}{a} - \frac{c}{a^2}\right)\left(1 - e^{-a\Delta t}\right) \qquad (22)$$

Where $\Delta s$ is the total displacement of the player during the time step $\Delta t$, $a = \frac{\beta}{m} + \frac{1}{\tau}$, $c = \frac{v_a}{\tau} + \frac{P}{m} - \frac{A}{m}e^{-|x|/b}$.

It is assumed that the time step is a random variable $\Delta t_{rand}$ drawn from a Gaussian distribution with a mean time step $\Delta t_m$ and a variance $\sigma_{\Delta t}$. Then we have to select an average direction along which the player displacement is $\Delta s$, calculated by Equation (22). Very important, at this point, is to make some reasonable assumptions in order to build an "objective function". The main criteria is to select a function correlated to the *strategy* of the player around which, in a necessarily *randomly* way, a *tactic* function should be added. The *strategy* depends on the player role. To carry out the numerical simulations discussed in this paper, a forward player was selected, whose "average" target is, obviously, to reach the goal. So it is straightforward to assume that the line direction, joining the players position and a point related to the goal, could be the main "*strategy objective function*" around which a *random angle* $\theta_{rand}$, expression of the "*tactic objective function*", influencing the direction selected by the player until a next interaction, could be introduced. Also it is reasonable to assume that the action of the player would be attracted by a position far from the border line. As a consequence the main selected direction has been *shifted* by a parametric angle: $\theta_{mean}= \theta_{goal}\pm\theta_{sh}$. The random variable $\theta_{rand}$ is supposed to belong to a Gaussian distribution characterized by $\theta_{mean}$ and $\sigma_\theta$.

For all the numerical simulations discussed in this paper, the following expression for the selected *random variable* has been considered:

$$V\_rand = \mu + n \cdot \sigma \cdot G\_norm \tag{23}$$

Thus:

$$\Delta t_{rand} = \Delta t_m + n_{\Delta t} \cdot \sigma_{\Delta t} \cdot G\_norm \tag{24}$$

$$\theta_{rand} = \theta_m + n_\theta \cdot \sigma_\theta \cdot G\_norm \tag{25}$$

where $n_{\Delta t}$ and $n_\theta$ are two number indicating the total range of variability (in the simulations shown in this paper: $n_{\Delta t}=2$ and $n_\theta=3$). In this case a *normal* "G_norm" Gaussian distributed stochastic variable ($\mu=0$ and $\sigma=1$) can be provided by the Box and Muller [25] algorithm:

$$G\_norm = \sqrt{[-2.d0 \cdot \ln(Y1_{rand})]} \cdot \cos(2.d0 \cdot \pi \cdot Y2_{rand}) \tag{26}$$

where $Y1_{rand}$ and $Y2_{rand}$ are two *independent* uniformly distributed random variables.

This indicates that the intrinsic routine which generates random variable values related to the selected Compilator (RAND in Fortran 97) is "called". This subroutine has to be "called" twice in order to obtain the two independent (pseudo-random) variables. To introduce an "*equivalent force*", due essentially to *tactic* player reasoning, it is assumed that at each point of the field is associated a different variance $\sigma_\theta(x,y)$ of the $\theta_{rand}$ variable. It means that *a player tactic reasoning*, specific to the area in which he is located, considered as a *random perturbation*, is superimposed to *the player strategic reasoning*.

Thus in Equation (25), $\theta_m$ could be interpreted as *the strategic objective*, while $n_\theta \cdot \sigma_\theta \cdot G\_norm$ is the *tactics objective* associated to each point. A soccer field 105 m long and 68 m wide has been considered for the numerical experiments performed. For the following discussions, regarding *the strategic objective*, it is assumed:

$$\theta_m(x,y) = \theta_{goal}(x,y) - \lambda \cdot \frac{y(player)}{wide} \tag{27}$$

where $\theta_{goal}(x,y)$ is the angle of the direction linking the point P(x,y) to the centre of the goalpost, $0 \leq x(player) \leq length(=105m)$ and $-wide/2$ ($=-34m$) $\leq y(player) \leq wide/2$ ($=34m$) are the coordinates of the player, while $\lambda$ is a parametric value. As *a tactics objective*, the suitability of the following functions has been explored:

$$\sigma_\theta^a = \left[1 - \frac{y(\text{player})}{\text{wide}/2}\right] \cdot \left[1 - \frac{x(\text{player})}{\text{lenght}}\right] \cdot \alpha \quad \sigma_\theta^b = \left[1 - \frac{y(\text{player})}{\text{wide}/2}\right] \cdot \left[1 - e^{-\frac{x(\text{player})}{\text{lenght}}}\right] \cdot \alpha$$

$$\sigma_\theta^c = \left[1 - e^{-\frac{y(\text{player})}{\text{wide}/2}}\right] \cdot \left[1 - e^{-\frac{x(\text{player})}{\text{lenght}}}\right] \cdot \alpha \quad \sigma_\theta^d = \left[1 - e^{-\frac{|y(\text{player})|}{\text{wide}/2}}\right] \cdot \left[1 - e^{-\frac{x(\text{player})}{\text{lenght}}}\right] \cdot \alpha$$

where α is a parametric angle. The initial position of the forward selected player was assumed to be P(30,30). Then an *average time step* between an interaction to the next one equal to $\Delta t_m$=3 sec is considered. In all the simulations discussed in this paper, the numerical values of Equation (21) and (22) have been chosen in such a way to assure that the average velocity is 2.2 m/s. Further a maximum of 30 steps have been considered. Lateral boundary lines have been assumed to be *reflective* without throw in, while if a goal or a corner foul have been occurred, the number of steps is automatically lowered. No *tactics* and just *strategy* is assumed at start, implying that $\sigma_\theta$=0.

In Figure 7 some numerical realizations related to only *strategy objective* are reported. The curvature of the path increases with the λ parameter as from Equation (27). Additionally, the path smoothness decreases with the time step variance, (Figure 7c), and the path trend shows a focalization toward x=0 and y=0.

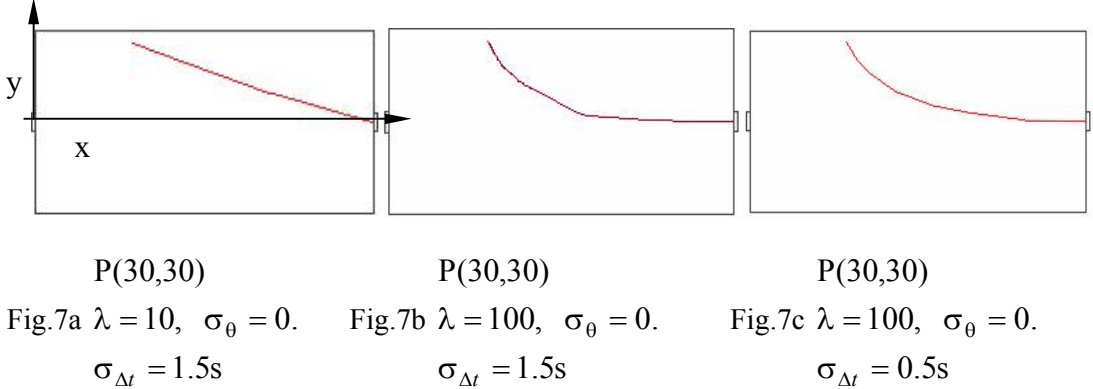

| P(30,30) | P(30,30) | P(30,30) |
|---|---|---|
| Fig.7a $\lambda = 10$, $\sigma_\theta = 0$. | Fig.7b $\lambda = 100$, $\sigma_\theta = 0$. | Fig.7c $\lambda = 100$, $\sigma_\theta = 0$. |
| $\sigma_{\Delta t} = 1.5s$ | $\sigma_{\Delta t} = 1.5s$ | $\sigma_{\Delta t} = 0.5s$ |

Figure 7: Numerical realizations related to the sole *strategy objective*.

Figure 8 shows two simulations including the *tactics objective* as well superimposed to a *strategy objective*. It is worth to note that $\sigma_\theta^a$ has been considered.

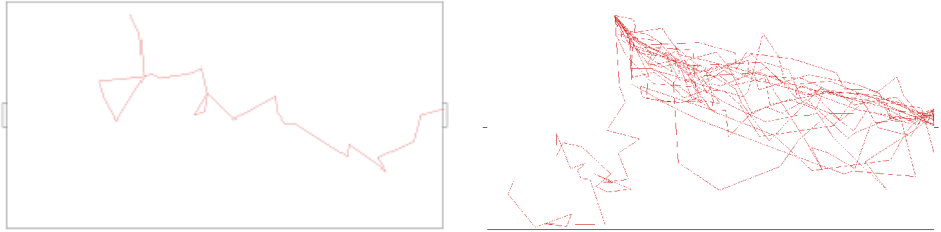

|  |  |
|---|---|
| P(30,30) | P(30,30) |
| Fig.8a  $\lambda = 100$, $\sigma_\theta^a = 50$. | Fig.8b  $\lambda = 100$, $\sigma_\theta^a = 50$. |
| $\sigma_{\Delta t} = 1.5$s;  1  *path* | $\sigma_{\Delta t} = 1.5$s;  2*0*  *path* |

Figure 8: Simulation outcomes including the *tactics objective*.

In Figure 9 *two different statistical realizations* of *equal parameters simulations* are shown. From the above figures it is easy to infer a strong predominance of the *strategy* on *tactics*.

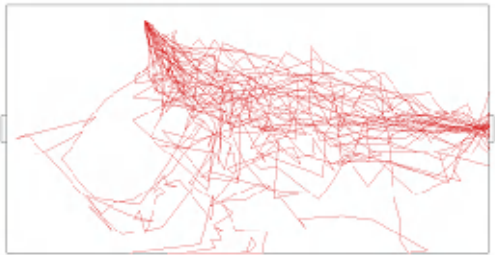 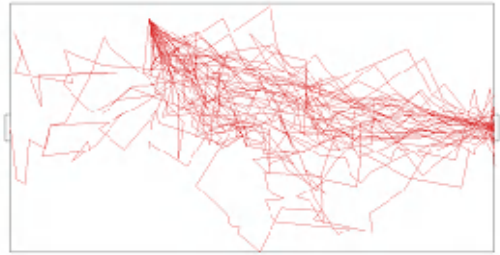

|  |  |
|---|---|
| P(30,30) | P(30,30) |
| Fig.9a  $\lambda = 100$, $\sigma_\theta^a = 50$. | Fig.9b  $\lambda = 100$, $\sigma_\theta^a = 50$. |
| $\sigma_{\Delta t} = 1.5$s;  4*0*  *path* | $\sigma_{\Delta t} = 1.5$s;  4*0*  *path* |

Figure 9: *Statistical realizations* of *equal parameters simulations*

In Figure 10 two different realizations of *equal parameters simulations* are shown, in which the *tactics* is prevalent on the *strategy*. This is due to the high value of the variance $\sigma_\theta^b$.

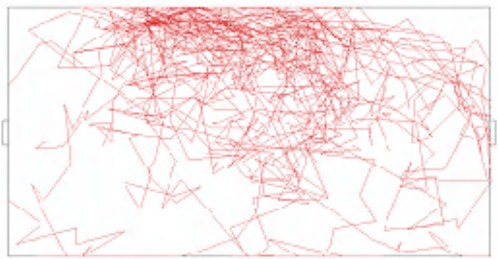 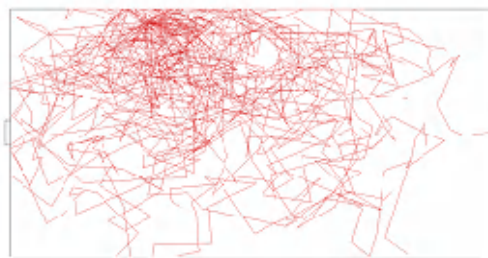

|  |  |
|---|---|
| P(30,30) | P(30,30) |
| Fig.10a  $\lambda = 10$, $\sigma_\theta^b = 400$. | Fig.10b  $\lambda = 10$, $\sigma_\theta^b = 400$. |
| $\sigma_{\Delta t} = 1.5$s;  4*0*  *path* | $\sigma_{\Delta t} = 1.5$s;  4*0*  *path* |

Figure 10: Two different realizations of *equal parameters simulations*

In Figure 11 the initial player position has been changed and two realization have been reported. It is interesting to observe that in this case, despite the fact that the variance associated to tactics is the same as in the previous case, the balance with the *strategy* is different, since the player initially is closer to the goal than he was in the previous case, so the strategy becomes stronger.

Figures 12, 13, 14 are confronted with Figure 11. The different path distributions is due to the different tactics objective selected for the simulations, respectively $\sigma_\theta^a$ $\sigma_\theta^c$ $\sigma_\theta^d$ with $\sigma_\theta^b$.

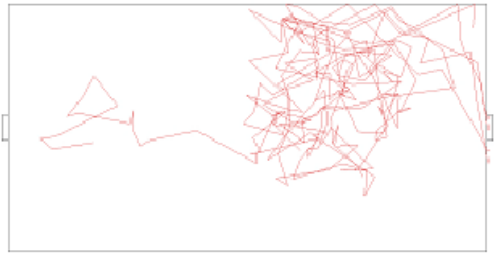 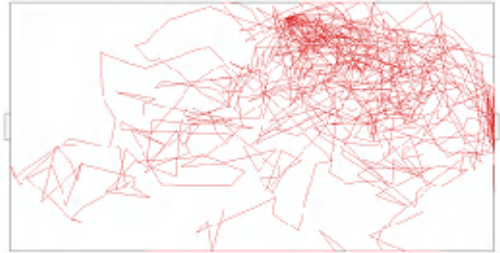

P(60,30)          P(60,30)

Fig.11a $\lambda = 10$, $\sigma_\theta^b = 400$.     Fig.11b $\lambda = 10$, $\sigma_\theta^b = 400$.

$\sigma_{\Delta t} = 1.5s$; $40\ path$       $\sigma_{\Delta t} = 1.5s$; $40\ path$

Figure 11: Two realization with the initial player position changed

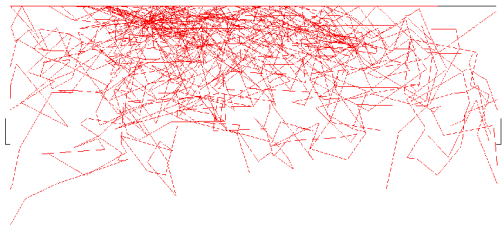 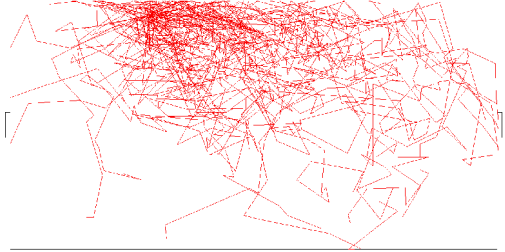

P(30,30)          P(30,30)

Fig.12a $\lambda = 10$, $\sigma_\theta^a = 400$.     Fig.12b $\lambda = 10$, $\sigma_\theta^a = 400$.

$\sigma_{\Delta t} = 1.5s$; $40\ path$       $\sigma_{\Delta t} = 1.5s$; $40\ path$

Figure 12: Two realization with the initial player position changed

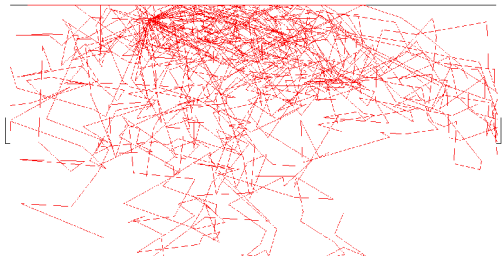 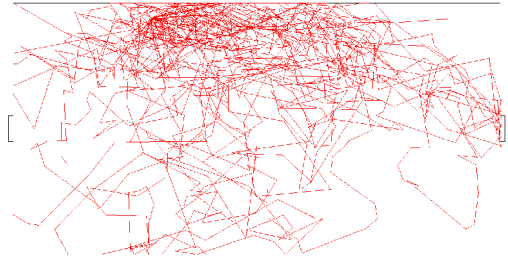

P(30,30)  P(30,30)

Fig.13a  $\lambda = 10$, $\sigma_\theta^c = 400$.  Fig.13b  $\lambda = 10$, $\sigma_\theta^c = 400$.
$\sigma_{\Delta t} = 1.5$s;  $40$  *path*  $\sigma_{\Delta t} = 1.5$s;  $40$  *path*

Figure 13: Two realization with the initial player position changed

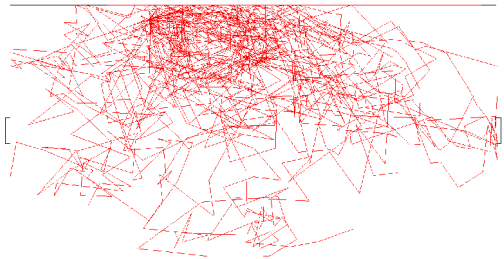 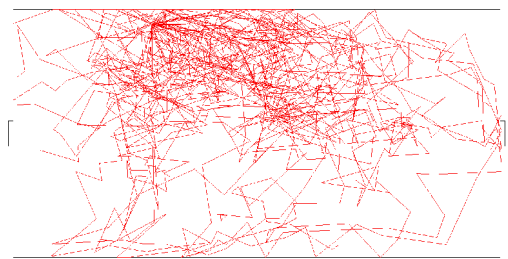

P(30,30)  P(30,30)

Fig.14a  $\lambda = 10$, $\sigma_\theta^d = 400$.  Fig.14b  $\lambda = 10$, $\sigma_\theta^d = 400$.
$\sigma_{\Delta t} = 1.5$s;  $40$  *path*  $\sigma_{\Delta t} = 1.5$s;  $40$  *path*

Figure 14: Two realization with the initial player position changed

From the above simulations, through a selection of reasonable assumptions, an evident Brownian behaviour of multiple paths related to a virtual player or to the ball seems to emerge clearly. Furthermore it is worthwhile to observe that there are some structures similarities between the experimental path reported in Figure 6 compared with Figure 8a. Then in particular, for multiple paths, the comparison between Figure 5 and Figures 10, 12, 13, 14 is very interesting.

As it was shown above, a particular path ensemble depends on the balance between the *strategy* and the *random tactics*. This observation demands an effort to capture, in the actual paths of actual players, a specific *strategy-tattics* behaviour which should be a *distinctive character* of each team sport athlete.

# 3  Conclusions

Starting from the well founded scientific knowledge that all the self organizing complex systems, especially the biological ones like the human body structure, from the DNA over to the heart or breath rate variability, are better described by non linear evolutions equations, the way these systems appear in their static, kinematics and dynamics forms (fractals) can be shown.

The connection among these different self organizing complex systems is given by the generalized Brownian Motion in every its known formulation: classic, skew, fractional, active and so on.

It is in the authors opinion, that Brownian dynamics could be assessed as one of the basic modelling of the mathematical alphabets of Life. On the basis of this knowledge, more remarkable, the usefulness of Brownian modelling is established, not only at a microscopic level or at a galactic level, as already established by Chandrasekhar [26], but at a macroscopic level as well, in which Brownian dynamics can describe the variability in the stride interval in walking, in running training or, as shown, in the modeling of fight in dual sport.

In this paper this method of modeling is extended to the analysis of the team sport motion during competition. However, if the focus is the study of team sport, it is better to implement a more advanced kind of Brownian motion, such as Fractional Brownian motion, Active Brownian motion, etc. In the approximation applied, it is found that the Active Brownian Motion with internal energy depots, is the most useful model that could be utilized in team motion modelling.

Furthermore, one special equation is proposed for the first time in the "Individual based" approach, describing the athletes motion in team sports, along with a numerical simulation of the trajectories. The motion paths obtained from the computational approach have been compared using experimental data of real games obtained by motion analysis systems. Normally active self driven motion can be found on different scales, starting from simple cells and over to higher organisations, such as the athletes human movements.

This kind of modelling is broad-spectrum, ranging from traffic related motion to organic micro-system motion, over to macroscopic human motion, and helps to validate the previous affirmation of the authors:

*Brownian dynamics is one of the basic modelling of the mathematical alphabets of Life.*